\documentclass[12pt]{revtex4}
\usepackage[dvips]{graphicx}
\begin{document}

\title{Tunable all-optical switching in periodic structures with liquid-crystal defects}

\author{Andrey E. Miroshnichenko, Igor Pinkevych, and Yuri S. Kivshar}

\address{Nonlinear Physics Centre and Centre for Ultra-high
bandwidth Devices for Optical Systems (CUDOS), Research School of
Physical Sciences and Engineering, Australian National University,
Canberra ACT 0200, Australia}


\begin{abstract}
We suggest that tunable orientational nonlinearity of nematic liquid
crystals can be employed for all-optical switching in
periodic photonic structures with liquid-crystal defects. We
consider a one-dimensional periodic structure of Si layers with a local defect
created by infiltrating a liquid crystal into a pore, and
demonstrate, by solving numerically a system of coupled nonlinear
equations for the nematic director and the propagating electric
field, that the light-induced Freedericksz transition can lead to a
sharp switching and diode operation in the photonic devices.
\end{abstract}

\maketitle

\section{Introduction}

During the past decade, photonic crystals (artificially fabricated
one-, two- and three-dimensional periodic dielectric materials)
have attracted a great deal of interest due to their ability to
inhibit the propagation of light over special regions known as
photonic band gaps~\cite{book}. Such photonic bandgap materials
are expected to revolutionize integrated optics and
micro-photonics due to an efficient control of the electromagnetic
radiation they provide, in such a way as semiconductors control
the behavior of the electrons~\cite{eli}.

In general, the transmission of light through photonic crystals
depends on the geometry and the index of refraction of the
dielectric material. Tunability of the photonic bandgap structures
is a key feature required for the dynamical control of light
transmission and various realistic applications of the photonic
crystal devices. One of the most attractive and practical schemes
for tuning the band gap in photonic crystals was proposed by Busch
and John~\cite{busch}, who suggested that coating the surface of an
inverse opal structure with a liquid crystal could be used to
continuously tune the band gap, as was confirmed later in
experiment~\cite{yoshino}. This original concept generated a stream
of interesting suggestions for tunable photonic devices based on the
use of liquid crystals infiltrated into the pores of a bandgap
structure~\cite{busch2}. The main idea behind all those studies is
the ability to continuously tune the bandgap spectrum of a periodic
dielectric structure using the temperature dependent refractive
index of a liquid crystal~\cite{yoshino,busch2,schuller,bjarklev},
or its property to change the refractive index under the action of
an applied electric field~\cite{zakhidov,escuti,khoo}.

Another idea of the use of liquid crystals for tunability of
photonic crystals is based on infiltration of individual
pores~\cite{mingaleev} and creation of liquid crystal
defects~\cite{villar,ozaki,greece}, and even defect-induced
waveguide circuits~\cite{mingaleev}. In this case, the transmission
properties can be controlled, for example, by tuning resonant
reflections associated with the Fano
resonances~\cite{fan,miroshnichenko} observed when the frequency of
the incoming wave coincides with the frequency of the defect mode.
As a result, the defect mode becomes highly excited at the frequency
of the resonant reflection, and it can be tuned externally, again by
an electric field or temperature.

However, liquid crystals  by themselves demonstrate a rich variety
of nonlinear phenomena (see, for example,
Refs.~\cite{zeldovich81,khoo81,ong83,zeldovich86}). Therefore,
nonlinear response of liquid crystals can be employed for
all-optical control of light propagation in periodic structures and
tunability of photonic crystals. In this paper, for the first time
to our knowledge, we analyze the possibility of {\em tunable
all-optical switching}  in one-dimensional periodic structure with a
liquid crystal defect. We demonstrate that the light field with the
intensity above a certain critical value corresponding to the
optical Freedericksz transition changes the optical properties of
the liquid-crystal defect such that the nonlinear transmission of
the photonic structure allows for all-optical switching, and the
similar concept can be employed for creating of a tunable
all-optical diode.

\section{Nonlinear transmission of a liquid crystal slab}

First, we study the light transmission of a single slab of nematic
liquid crystal and derive a system of coupled nonlinear equations
for the liquid-crystal director reorientation in the presence of the
propagating electric field of a finite amplitude. The corresponding
steady-state equation for the director $\mathbf{n}$ can be obtained
by minimizing the free energy which can be written in the following
form~\cite{zeldovich81,degennes}
\begin{equation}
\label{eq:free_elastic1}
\begin{array}{l}
{\displaystyle f=f_{\rm el}+f_{\rm opt},}\\*[9pt]
{\displaystyle
f_{\rm
el}=\frac{1}{2}\left[K_{11}(\nabla\cdot\mathbf{n})^2+K_{22}(\mathbf{n}\cdot\nabla\times\mathbf{n})^2\;
    +K_{33}(\mathbf{n}\times\nabla\times\mathbf{n})^2\right],}\\*[9pt]
{\displaystyle f_{\rm opt}=-(1/16\pi)\mathbf{D\cdot E^*}},
\end{array}
\end{equation}
where $f_{\rm el}$ is the elastic part and $f_{\rm opt}$ is the
optical part of the energy density. Here $K_{11}$, $K_{22}$ and
$K_{33}$ are splay, twist, and bend elastic constants, respectively,
$\mathbf{D}= \hat{\epsilon} \mathbf{E}$, $\hat{\epsilon}$ is the
dielectric tensor, and the real electric field is taken in the form
$\mathbf{E}_{real}=(1/2)[\mathbf{E}(\mathbf{r})\exp(-i\omega
t)+\mathbf{E}^{*}(\mathbf{r})\exp(i\omega t)]$.

\begin{figure}[h]
\centering
\includegraphics[width=120mm]{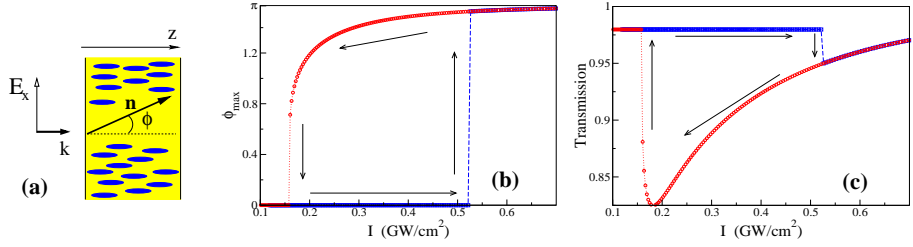}
\caption{Nonlinear transmission of a liquid-crystal slab. (a)
Schematic of the problem. (b,c) Maximum angle of the director and
transmission vs. the light intensity in the slab. Blue and red
curves correspond to the increasing and decreasing light intensity,
respectively.}
  \label{fig0}
\end{figure}

We assume that linearly polarized light wave propagates normally to
the liquid-crystal slab with the initial homeotropic director
orientation along $z$ [see Fig.~\ref{fig0}(a)]. Under the action of
the electric field polarized outside the slab along $x$, the
director can change its direction in the $(x,z)$ plane and,
therefore, we write the vector components of the director in the
form $\mathbf{n}=\{\sin\phi(z),0,\cos\phi(z)\}$. Then the elastic
part of the free energy density can be written as
\begin{eqnarray}
\label{eq:free_elastic2}
    f_{\rm el}=\frac{1}{2}\left(K_{11}\sin^2\phi+K_{33}\cos^2\phi\right)\left(\frac{d
\phi}{d z}\right)^2\;.
\end{eqnarray}
Taking into account that the dielectric tensor $\hat{\epsilon}$ can
be expressed in terms of the director components,
$\epsilon_{ij}=\epsilon_{\bot}\delta_{ij}+\epsilon_a n_in_j$, where
$\epsilon_a=\epsilon_{||}-\epsilon_{\bot}$ and $\epsilon_{||}$,
$\epsilon_{\bot}$ are the liquid crystal dielectric constants at the
director parallel and perpendicular to the electric vector, we can
write
\begin{eqnarray}\label{eq:tensor}
   \hat{ \epsilon}=\left(
    \begin{array}{ccc}
     \epsilon_{\bot}+\epsilon_a\sin^2\phi&0&\epsilon_a\sin\phi\cos\phi\\
     0&\epsilon_{\bot}&0\\
     \epsilon_a\sin\phi\cos\phi&0&\epsilon_{\bot}+\epsilon_a\cos^2\phi
    \end{array}
    \right)\;.
\end{eqnarray}
As a result, the optical part of the free energy density takes the
form
\[f_{\rm
opt}=-\frac{\epsilon_a}{16\pi}\left[\sin^2\phi|E_x|^2+\cos^2\phi|E_z|+
+\sin\phi\cos\phi(E_xE_z^*+E_zE_x^*)\right]-\frac{\epsilon_{\bot}}{16\pi}|\mathbf{E}|^2\;.
\]

After minimizing the free energy (\ref{eq:free_elastic1}) with
respect to the director angle $\phi$, we obtain the nonlinear
equation for the director in the presence of the light field

\begin{eqnarray}
\label{eq:director2} A(\phi) \frac{d^2\phi}{d
z^2}-B(\phi)\left(\frac{d\phi}{dz}\right)^2+
\frac{\epsilon_a\epsilon_{\bot}(\epsilon_a+\epsilon_{\bot})|E_x|^2\sin2\phi}{16\pi(\epsilon_{\bot}+\epsilon_a\cos^2\phi)^2}=0,
\end{eqnarray}
where $A(\phi)= (K_{11}\sin^2\phi+K_{33}\cos^2\phi)$,
$B(\phi)=(K_{33}-K_{11})\sin\phi\cos\phi$, and we take into account
that, as follows from $D_{\rm z}=0$, that the electric vector of the
light field has the longitudinal component,
$E_z=-(\epsilon_{xz}/\epsilon_{zz})E_x=-[\epsilon_a\sin\phi\cos\phi/(\epsilon_{\bot}+\epsilon_{a}\cos^2\phi)]E_x$
(see also Ref.~\cite{zeldovich81}).

From the Maxwell's equations, we obtain the equation for the
electric field $E_x$,
\begin{eqnarray}
\label{eq:full_set} \frac{d^2
E_x}{dz^2}+k^2\frac{\epsilon_{\bot}(\epsilon_{\bot}+\epsilon_a)}{\epsilon_{\bot}+\epsilon_a\cos^2\phi}E_x=0,
\end{eqnarray}
where $k=2\pi/(\lambda c)$. Moreover, it can be
shown~\cite{zeldovich81,ong83} that the $z$ component of the
Poynting vector $I=S_z=(c/8\pi)E_xH^{*}_y$ remains constant during
the light scattering and, therefore, it can be used to characterize
the nonlinear transmission results.

As the boundary conditions for the coupled nonlinear equations
(\ref{eq:director2}) and (\ref{eq:full_set}), we assume that there
is an infinitely rigid director anchoring at both surfaces of the
slab, i.e.
\begin{eqnarray}
\label{eq:boundaries1} \phi(0)=\phi(L)=0,
\end{eqnarray}
and also introduce the scattering amplitudes for the optical field
\begin{eqnarray}
\label{eq:boundaries2}
E_x(z)=\left\lbrace %
\begin{array}{lc}
    {\cal E}_{\rm in}\exp(ikz)+{\cal E}_{\rm ref}\exp(-ikz),& z\le 0,\\
    {\cal E}_{\rm out}\exp(ikz), & z\ge L,
   \end{array}
   \right.
   \end{eqnarray}
where $L$ is the thickness of the liquid-crystal slab, ${\cal
E}_{\rm in}$, ${\cal E}_{\rm ref}$, and ${\cal E}_{\rm out}$ are the
electric field amplitudes of incident, reflected, and outgoing
waves, respectively.

To solve this nonlinear problem, first we fix the amplitude of the
outgoing wave ${\cal E}_{\rm out}$ and find unique values for the
amplitudes of the incident, ${\cal E}_{\rm in}$, and reflected ,
${\cal E}_{\rm ref}$, waves. By using the so-called \textit{shooting
method}~\cite{nr}, in Eq.~(\ref{eq:director2}) for the director we
fix the amplitude of the outgoing wave and, assuming that
$\phi(L)=0$ at the right boundary, find the derivative
$(d\phi/dz)_{z=L}$ such that after integrating we obtain a vanishing
value of the director at the left boundary, i.e. $\phi(0)=0$.
Because Eq.~(\ref{eq:director2}) is a general type of the nonlinear
pendulum equation, we look for periodic solutions with the period
$2L$. Obviously, there exists an infinite number of such solutions
and, therefore, there is an infinite set of the derivatives
$(d\phi/dz)_{z=L}$ which satisfy Eq.~(\ref{eq:director2}) and the
condition (\ref{eq:boundaries1}). All such solutions correspond to
some extrema points of the free energy of the system. However, we
are interested only in that solution which realizes the minimum of
the free energy. By analyzing our coupled nonlinear equations in a
two-dimensional phase space, we can show that the corresponding
solution lies just below the separatrix curve, and it has no node
between the points $z=0$ and $z=L$. This observation allows us to
reduce significantly the domain for our search  for the required
values of the derivative $(d\phi/dz)_{z=L}$.

\begin{figure}[h]
  \centering
  \includegraphics[width=120mm]{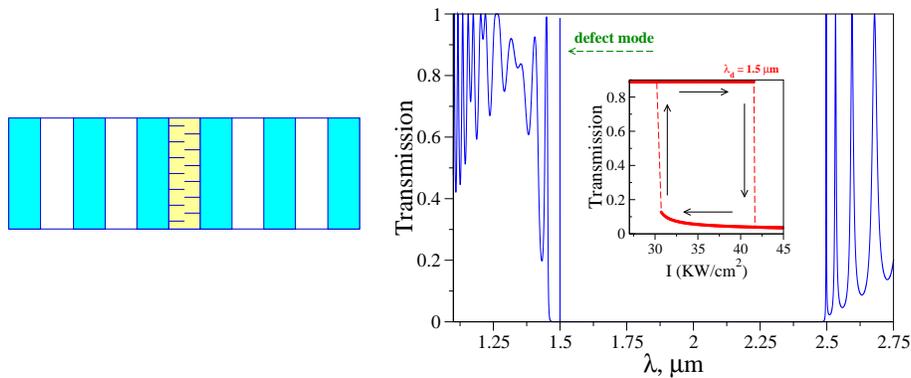}
  \caption{Transmission of an one-dimensional
periodic structure with an embedded liquid-crystal defect. In the
linear regime, the transmission is characterized by the presence of
an in-gap resonant peak due to the excitation of a defect mode.
Nonlinear transmission displays bistability at the defect-mode
frequency with two different thresholds for "up" and "down"
directions and a hysteresis loop (see the insert).}
  \label{fig1}
\end{figure}

The obtained solutions can be characterized by the maximum angle
$\phi_{\rm max}$ of the director deviation which, as is intuitively
clear, should be reached near or at the middle of the slab. In
Fig.~\ref{fig0}(b,c), we plot the maximum angle $\phi_{\rm max}$ and
the transmission coefficient of the liquid-crystal slab, defined as
$T=|{\cal E}_{\rm out}|^2/|{\cal E}_{\rm in}|^2$, vs. the light
intensity. For numerical calculations, we use the parameters
$K_{11}=4.5\times10^{-7}$ dyn, $K_{33}=9.5\times10^{-7}$ dyn,
$\epsilon_a=0.896$, $\epsilon_{\bot}=2.45$, $L=200 nm$, and
$\lambda=1.5\mu m$, that correspond to the PAA liquid
crystal~\cite{stephen}; because of a lack of the corresponding data
at the wavelength $\lambda=1.5\mu m$, the values of the dielectric
constant are taken from the optical range.

From the results presented in Fig.~\ref{fig0}(b,c), we observe sharp
jumps of the director maximum angle $\phi_{\rm max}$ and the
transmission coefficient $T$ due to the \textit{optical Freedericksz
transition} in the liquid-crystal defect. However, a variation of
the transmission coefficient during this process is not larger than
$15\%$. The threshold of the optical Freedericksz transition appears
to be different for the increasing and decreasing intensity of the
incoming light, so that this nonlinear system is bistable, and it
displays a hysteresis behavior. The bistable transmission of the
liquid-crystal slab is similar to that predicted for the slab of PAA
liquid crystal in the geometric optics approximation~\cite{ong83},
and such a behavior is explained by the existence of the metastable
state which the system occupies at the decreasing light
intensity~\cite{zeldovich81,ong83}.

\section{Liquid-crystal defect in a periodic photonic structure}

Now, we study the similar problem for a liquid-crystal defect
infiltrated into a pore of the periodic structure created by Si
layers with the refractive index $n=3.4$. For simplicity, we
consider a one-dimensional structure with the period $a=400 nm$ and
the layer thickness $d_1 = 200 nm$, that possesses a frequency gap
between $1.4 \mu m$ and $2.5 \mu m$. We assume that one of the holes
is infiltrated with a PAA nematic liquid crystal with
$\epsilon_{\bot}=2.45$. Such a defect modifies the linear
transmission of the periodic structure by creating a sharp
defect-mode peak at the wavelength $\lambda_d\approx1.5 \mu m$, as
shown in Fig.~\ref{fig1}.

\begin{figure}[h]
  \centering
  \includegraphics[width=118mm]{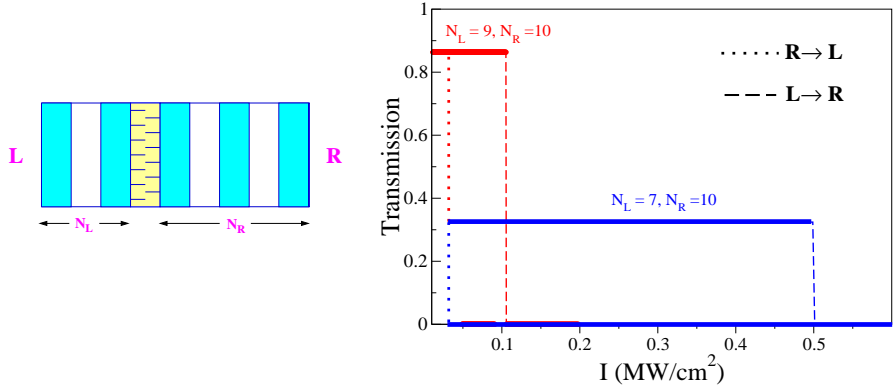}
  \caption{Example of a tunable all-optical diode based
on the optical Freedericksz transition in a liquid-crystal defect.
Asymmetrically placed defect leads to different threshold
intensities of the switching for the waves propagating from the
right and left, respectively.}
  \label{fig2}
\end{figure}

To solve the nonlinear transmission problem, we employ the transfer
matrix approach~\cite{yeh} implementing it for the solution of the
full system of coupled equations (\ref{eq:director2}) and
(\ref{eq:full_set}). By tuning the input intensity at the defect
mode, we observe the same scenario as for a single liquid-crystal
slab [cf. the insert in Fig.~\ref{fig1} and Fig.~\ref{fig0}(c)].
Namely, there exists a hysteresis loop in the transmission with two
different thresholds for the increasing and decreasing intensities.
The difference is, however, in the values. Due to a small width of
the defect-mode resonance, even a small reorientation of the
director leads to a sharp (up to $90\%$) change in the transmission.
Another significant difference is that the threshold values are
\textit{lower by four orders of the magnitude}, for a given periodic
structure for which we take 10 layers from each side of the defect.

Finally, we notice that in a finite periodic structure the defect
placed asymmetrically (see Fig.~\ref{fig2}) allows to create a
nonreciprocal device when the threshold intensities for the
molecular reorientation differ for the light propagating from the
right and left. 
This feature is associated with the operation of an
\textit{optical diode}~\cite{scalora,gallo}. As can be seen in
Fig.~\ref{fig2}, by shifting the infiltrated liquid-crystal defect
closer to one of the edges of the structure and fixing the total
length of the structure, we can increase the switching power and
extend the diode operation region decreasing the transmission power.
Moreover, these results show that the threshold intensities depend
strongly on the number of periods to the structure edge, due to a
stronger confinement of the defect mode. Also, this gives us a
possibility to reduce significantly the switching power simply by
taken larger number of periods in the photonic structure.

\section{Conclusions}

We have demonstrated that the orientational nonlinearity of nematic
liquid crystals can be employed to achieve tunable all-optical
switching and diode operation in periodic photonic structures with
infiltrated liquid-crystal defects. For the first time to our
knowledge, we have solved a coupled system of nonlinear equations
for the nematic director and the propagating electric field for the
model of a one-dimensional periodic structure created by Si layers
with a single (symmetric or asymmetric) pore infiltrated by a liquid
crystal. We have demonstrated that the threshold of the optical
Freedericksz transition in the liquid-crystal defect is reduced
dramatically due to multiple reflections in the periodic structure,
so that such a defect may allow a tunable switching and diode
operation in the photonic structure.

\section*{Acknowledgements}

The work has been supported by the Australian Research Council. The
authors thank B.Ya. Zeldovich, I.C. Khoo, M. Karpierz, and O.
Lavrentovich for useful discussion of our results and suggestions,
and I.V. Shadrivov for the help in numerical simulations.

\end{document}